\documentclass[aps,prl,twocolumn]{revtex4-1}
\usepackage[utf8]{inputenc} % If utf8 encoding
\usepackage[T1]{fontenc}    %
\usepackage[english]{babel} % English please
\usepackage{amssymb,amsmath,mathtools} % Math
\usepackage{graphicx} 
\usepackage{xcolor}

\usepackage{hyperref}

\newcommand{\erfc}{{\rm erfc}\,}
\newcommand{\Tr}{{\rm Tr}\,}

\begin{document}

\title{Universal 
Signature from Integrability to Chaos in 
%%%%%
Dissipative 
%%%%%
Open Quantum Systems}

\author{Gernot Akemann}\email{akemann@physik.uni-bielefeld.de}\affiliation{Faculty of Physics, Bielefeld University, Postfach 100131, 33501 Bielefeld, Germany,\\
Department of Mathematics, Royal Institute of Technology (KTH), 
Brinellv\"agen 8, 114 28 Stockholm, Sweden}
\author{Mario Kieburg}\email{m.kieburg@unimelb.edu.au}\affiliation{School of Mathematics and Statistics, University of Melbourne, 813 Swanston Street, Parkville, Melbourne VIC 3010, Australia}
\author{Adam Mielke}\email{amielke@math.uni-bielefeld.de}\affiliation{Faculty of Physics, Bielefeld University, Postfach 100131, 33501 Biele\-feld, Germany}\author{Toma\v{z} Prosen}\email{tomaz.prosen@fmf.uni-lj.si}\affiliation{Physics Department, Faculty of Mathematics and Physics, University of Ljubljana, Ljubljana 1000, Slovenia}

\date{\today}

\begin{abstract}

We study the transition between integrable and chaotic behaviour in dissipative open quantum systems, exemplified by a boundary driven quantum spin-chain. The repulsion between the complex eigenvalues of the corresponding Liouville operator in radial distance $s$ is used as a universal measure.
The corresponding level spacing distribution is well fitted
by that of a static two-dimensional Coulomb gas with harmonic potential at inverse temperature $\beta\in[0,2]$.
Here,  $\beta=0$ yields the two-dimensional Poisson distribution, matching the integrable limit of the system, and $\beta=2$ equals the distribution 
obtained from the complex Ginibre ensemble,
describing the fully chaotic limit. Our findings generalise the results of  Grobe, Haake and Sommers who derived a universal cubic level repulsion for small spacings $s$.  We collect mathematical evidence for the universality of the full level spacing distribution in the fully chaotic limit at $\beta=2$. It holds for all three Ginibre ensembles {of} random matrices with independent real, complex or quaternion matrix elements. 

\end{abstract}
 
\date{\today}
\maketitle

%%%%%%%%%%%%%%%%%%%%%%%%%%%%%%
\paragraph{Introduction.}

It has been a long discussed question
how classically integrable and chaotic behaviour carries over to the quantised world.
A simple spectral measure was found in the spacing between neighbouring eigenvalues of the corresponding Hamiltonian $H$. 
For closed systems it is Hermitian, $H=H^\dag$, with real eigenvalues.
Berry and Tabor
conjectured~\cite{BT-conj} for quantum integrable systems to generically follow the one-dimensional (1D)
Poisson distribution $p_{\rm P}^{\rm (1D)}(s)=e^{-s}$. In contrast, 
Bohigas, Giannoni and Schmit (BGS) conjectured~\cite{BGS-conj} {(cf. \cite{CGV})}
chaotic systems~\cite{K-chaotic} to follow random matrix theory (RMT) statistics in the corresponding symmetry class. 
Initially Dyson ~\cite{Dyson} had offered a first classification within RMT, distinguishing systems without or with time-reversal at (half-)integer spin which is the celebrated ''threefold way''.
Much evidence has been given to support this spectral classification in quantum systems, including neutron scattering, quantum billiards \cite{BGS-conj}, or the hydrogen atom in a magnetic field \cite{Wintgen} to name a few, cf. \cite{GMGW,Stoeckmann} for standard references. 
Starting from Berry's diagonal approximation \cite{Berry}
the BGS-conjecture is now well understood from a semi-classical expansion \cite{SR,Haake}.

Non-Hermitian operators play an equally important role in physics, e.g. in disordered systems \cite{HatanoNelson} or Quantum Chromodynamics (QCD) with chemical potential \cite{Misha}. 
Shortly after BGS, the above spectral distinction between integrable and chaotic was extended by Grobe, Haake and Sommers (GHS) \cite{GHS1988} to Markovian dissipative open quantum systems. These follow a Lindblad master equation  
\begin{equation}\label{Lindblad}
\frac{d \rho}{dt}(t)={L}\rho(t)\ ,
\end{equation}
with ${L}$ the Liouville and $\rho$ the density operator, cf. \cite{Breuer}. 
Postponing a detailed discussion 
for our example of a quantum XXZ spin-chain, see~\cite{Prozen2011,BP2012}, the eigenvalues of ${L}$ are real or come in complex conjugate pairs and can be used to characterise integrable or chaotic behaviour, see below. 
Indeed this has been observed in many examples for dissipative chaotic systems~\cite{bookHaake},
for the QCD Dirac operator with chemical potential~\cite{Wettig}, the adjacency matrix of directed graphs~\cite{YQWG2015} {and hard-core bosons with asymmetric hopping on a one-dimensional lattice 
at weak disorder \cite{H1}}.
In \cite{GHS1988} GHS studied periodically kicked tops with damping  and the corresponding discrete quantum map. 
In the integrable limit 
they found agreement between 
the nearest neighbour spacing in radial distance $s$  of its complex bulk eigenvalues
and the two-dimensional (2D) Poisson distribution 
\begin{equation}\label{Poisson-2d}
p_{\rm P}^{\rm (2D)}(s)= \frac{\pi}{2} s e^{-\pi s^2/4} ,
\end{equation}
which are local quantities.
In the fully chaotic limit {the spacing distribution agrees with the corresponding} distribution of the Ginibre ensemble \cite{Ginibre} of complex Gaussian non-Hermitian random matrices (GinUE), given by
\cite{GHS1988}
\begin{equation}\label{GinibreSpacing}
p_{\rm GinUE}(s) =
\left(\prod_{k=1}^{\infty}\frac{\Gamma(1 + k, s^2)}{k!}\right)
\sum_{j=1}^{\infty} \frac{2s^{2j+1}e^{-s^2}}{\Gamma(1 + j, s^2)},
\end{equation}
with $\Gamma(1 + k, s^2)=\int_{s^2}^\infty t^k e^{-t}dt$ the incomplete Gamma function. GHS conjectured that the local spectrum  of a generic chaotic 
%%%%%%%%
dissipative
%%%%%%%%
open quantum system  in the bulk should follow the same statistics. 
This was somewhat surprising, as they showed that the complex Ginibre ensemble leading to \eqref{GinibreSpacing} does not satisfy the global symmetries of  dissipative open quantum systems~\cite{GHS1988}, unlike its real counterpart. 
They showed in 
\cite{GH1989} that based on these symmetries, using perturbative arguments
for small distance $s$ the repulsion is universally cubic.
This repulsion is shared 
%though 
by the complex Ginibre ensemble~\eqref{GinibreSpacing}, as well as by a larger class of complex normal random matrices \cite{Oas}. {The global statistics of Lindblad operators has also been compared to random matrices, cf. \cite{Karol, Can1,Can2,TP2}.}

Our goals are, first, to provide a further example for the GHS conjecture for complex spectra of integrable or quantum chaotic systems 
to be true,
%%%% 
given by boundary driven quantum spin-chains. These are many-body systems with no meaningful semi-classical limit, so the term {\em quantum chaos} is
understood as absence of integrability 
%structure 
or weak coupling thereof while its rigorous definition is still lacking.
%%%%
Second, we will show that in the intermediate regime
the full spacing distribution  is very well described by a static 2D Coulomb gas at inverse temperature $\beta\in(0,2]$ in a harmonic potential. Its joint distribution of the set $z$ of $N$ point charges at 
%%%%%%%
rescaled positions 
$\sqrt{{2}/{\beta}}\ z_{i=1,\ldots,N}\in\mathbb{C}$ 
\cite{betalim}
reads \cite{Peterbook}
\begin{equation}
\label{2DCoulomb}
\mathcal{P}_{\beta}(z) \propto\ \exp\left[-{\sum}_{i=1}^N|z_i|^2+\frac{\beta}{2}{\sum}_{i\neq j}^N\ln|z_i-z_j|\right].
\end{equation}
For $\beta=0$ this leads to the Poisson distribution \eqref{Poisson-2d} \cite{bookHaake}, whereas $\beta=2$ corresponds to the level spacing distribution~\eqref{GinibreSpacing} \cite{GHS1988}. 
Third, we collect mathematical evidence for  the fully chaotic case \eqref{GinibreSpacing} at $\beta=2$
to be universal in the bulk of the spectrum, regardless of the constraints \cite{GHS1988}. With bulk we mean to stay macroscopically away from any edge or critical points (here the real line) of the spectrum.
This universality holds for  the complex, real~ \cite{BorodinSinclair} and quaternion Ginibre ensemble - to be presented here - and for non-Gaussian extensions~\cite{TaoVu} of the two former.
This is in contrast to random matrices with real spectra, where quantum chaotic behaviour is distinct for the three Dyson classes, corresponding to a 1D log-gas at different values $\beta=1,2,4$. For complex bulk eigenvalues of chaotic systems the possibility to distinguish their global symmetry is thus lost.
To prepare our 2D data from the Liouville operator ${L}$ for a comparison we need to unfold the complex spectrum. While this is straightforward for  real spectra~\cite{GMGW},
we discuss the literature \cite{Wettig} and present our method below.\\

\paragraph{Integrable and Non-Integrable 
Quantum Spin-Chains.}

The system we consider is a Heisenberg XXZ Hamiltonian $H$ of $N$ spins $1/2$, comprising nearest and next-to-nearest neighbour interactions,
\begin{equation}\label{Hamiltonian}
\begin{split}
H=&J{\sum}_{l=1}^{N-1}(\sigma_l^x\sigma_{l+1}^x+\sigma_l^y\sigma_{l+1}^y+\Delta \sigma_l^z\sigma_{l+1}^z)\\
&+
J'{\sum}_{l=1}^{N-2}(\sigma_l^x\sigma_{l+2}^x+\sigma_l^y\sigma_{l+2}^y+\Delta' \sigma_l^z\sigma_{l+2}^z)\ ,
\end{split}
\end{equation}
with $J,J',\Delta,\Delta'\in\mathbb{R}$. We denote the three Pauli matrices by $\sigma_l^\alpha$, $\alpha=x,y,z$, for each single spin $l=1,\ldots,N$. To each spin a dephasing operator
\begin{equation}\label{dephasing-spin}
L_l=\sqrt{\gamma}\sigma_l^z,\ l=1,\ldots, N\ {\rm and}\ \gamma>0
\end{equation}
is associated. Additionally, we introduce {dissipation of polarisation} at the two ends of the spin-chain via the Lindblad operators
\begin{equation}\label{dephasing-boundary}
\begin{split}
L_{-1}=&\sqrt{\gamma_{\rm L}^+}\sigma_1^+,\ L_{0}=\sqrt{\gamma_{\rm L}^-}\sigma_1^-,\\
 L_{N+1}=&\sqrt{\gamma_{\rm R}^+}\sigma_N^+,\ L_{N+2}=\sqrt{\gamma_{\rm R}^-}\sigma_N^-,
 \end{split}
\end{equation}
where $\gamma_{\rm L}^\pm,\gamma_{\rm R}^\pm>0$ and $\sigma_l^\pm=\sigma_l^x\pm i\sigma_l^y$. The Liouville operator ${L}$ acting on a density operator $\rho$ in the master equation \eqref{Lindblad} is given by~\cite{Prozen2011,BP2012}
\begin{equation}\label{Liouvillian}
{L}\rho=-i[H,\rho]+{\sum}_{l=-1}^{N+2}(2L_l\rho L_l^\dagger-\{L_l^\dagger L_l,\rho\}).
\end{equation}
The commutator and anti-commutator are denoted by $[.,.]$ and $\{.,.\}$, respectively, cf.  \cite{Breuer}. 

What we are interested in is the spectral statistics of the Liouville operator ${L}$ considered as a $(4^N-1)\times(4^N-1)$ real matrix, acting on the vector space of density operators. The reduction in dimension by one results from the fixed trace condition on $\rho$ and is represented by the identity matrix. The operator ${L}$ is real because $\rho\to {L}\rho$ preserves the Hermiticity.   The statistics of ${L}$ should indicate whether the Lindblad master equation \eqref{Lindblad}
behaves in an integrable or chaotic way. For this purpose we recall some properties of the operator ${L}$ in our example.

Switching off 
%the 
all 
incoherent processes $\gamma=\gamma_{\rm L}^\pm=\gamma_{\rm R}^\pm=0$, the operator  ${L}$ becomes a real anti-symmetric {(because of 
$\Tr\rho_1[H,\rho_2]=-\Tr[H,\rho_1]\rho_2$)} and chiral (due to $[H,\rho]^T=-[H,\rho^T]$) matrix, so that the 
%1D 
%%%%%%
spectrum 
becomes 1D and 
%%%%%%
is purely imaginary and symmetric about the origin. When also suppressing the next-to-nearest neighbour interactions ($J'=0$) the spectrum 
%becomes 
is 
completely integrable. With increasing $J'\neq0$ chaotic behaviour will take over and 
%one would expect 
Wigner's $\beta=1$ statistics in the bulk of the spectrum
%%%%%%%%%%
applies, see \cite{DAlessio} for a review of the standard 1D RMT analysis of this setup.
%%%%%%%%%

The situation changes drastically when the {dissipative processes are} switched on ($\gamma,\gamma_{\rm L}^\pm,\gamma_{\rm R}^\pm\neq0$). Then, the Liouville operator  ${L}$ becomes a real non-symmetric matrix and its eigenvalues spread into the complex plane. Nonetheless, there is still a good quantum number which has to be taken into account, namely the total spin polarisation $S=\sum_{l=1}^N\sigma_l^z$. It keeps the coherent processes invariant due to $[H,S]=0${, while all additional incoherent dissipative processes result in} the following weak symmetry of the Liouvillian~\cite{BP2012}
\begin{equation}\label{Weak-sym}
[{L}(\rho),S]={L}([\rho,S]),
\end{equation}
{which} is equivalent to the vanishing commutator of the 
matrix representations of ${L}$ and of $[S,.]$.

Let $|s,n\rangle$ be an eigenstate of $H$ with $S|s,n\rangle=s|s,n\rangle$ and $s=-N/2,-(N-2)/2,\ldots,N/2$. Then, the eigenvalue equation of the state $|s,n\rangle\langle s',n'|$ under the adjoint action of $S$ is
\begin{equation}
[S,|s,n\rangle\langle s',n'|]=(s-s')|s,n\rangle\langle s',n'|.
\end{equation}
Defining $M=N-s+s'\in\{0,1,\ldots,2N\}$, the dimension 
%%%%%%%%%%%%%
$\kappa$ 
%%%%%%%%%%%%%
of the eigenspace of the fixed quantum number $s-s'=N-M$ is given by 
%%%%%%%%%%%%%
$\kappa=\binom{2N}{M}-\delta_{NM}$, 
%%%%%%%%%%%%%%
where the Kronecker delta represents the 
identity matrix which obviously belongs to the $M=N$ state space. Therefore, ${L}$ decomposes into block matrices and one needs to study the spectral statistics of each of these matrices separately. Since we are interested in a good statistical error, it is favourable to choose $M$ close to $N$,
%%%%%%%%%%%%%
as then the number of eigenvalues $\kappa\sim 2^{2N}/N$ grows exponentially fast for large $N$.
%%%%%%%%%%%%%
\\

\paragraph{Comparing Data with Predictions.}
	\begin{figure}[t!]
		\centering
		\includegraphics[width=0.49\linewidth,angle=0]{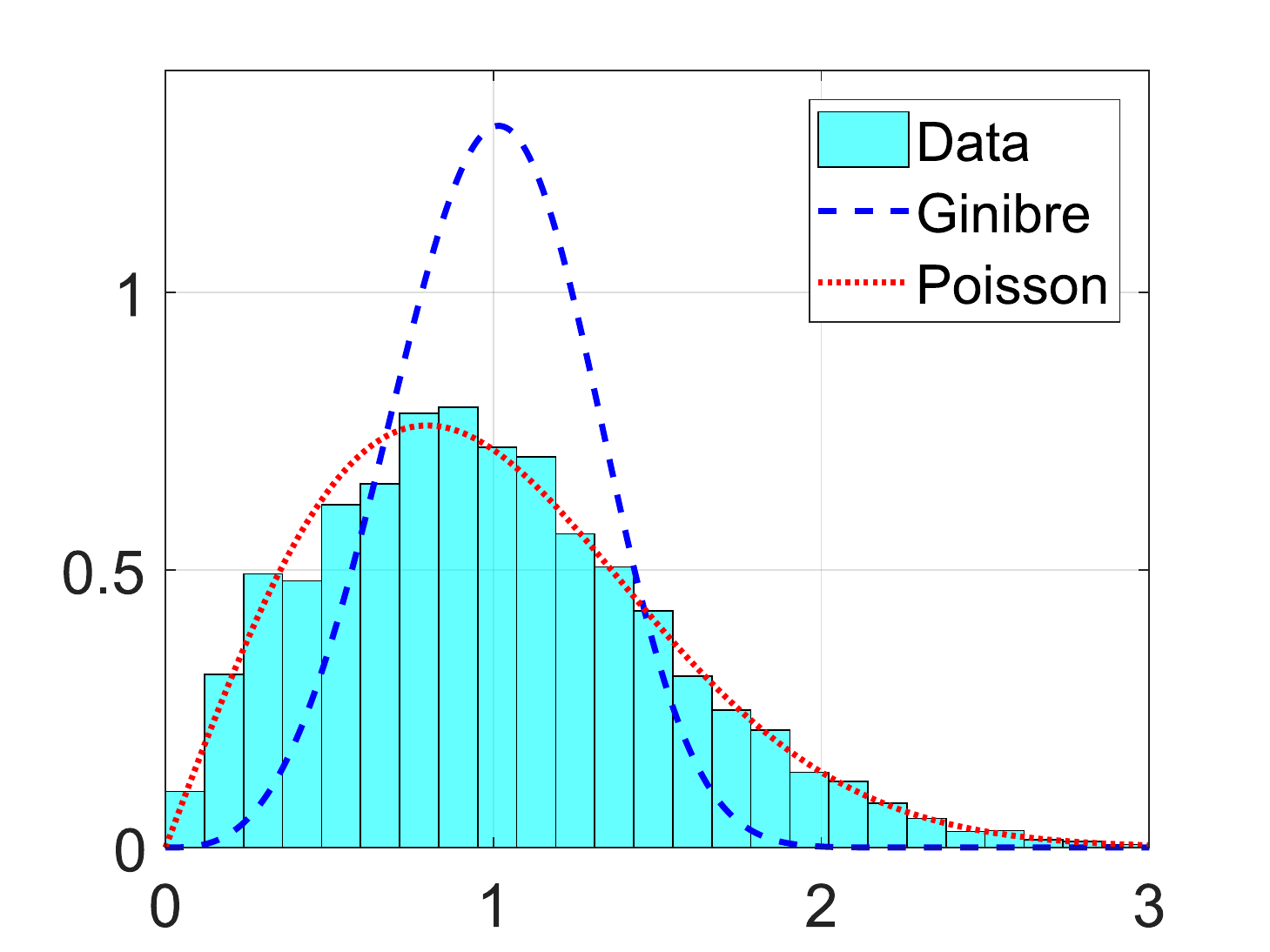}
		\includegraphics[width=0.49\linewidth,angle=0]{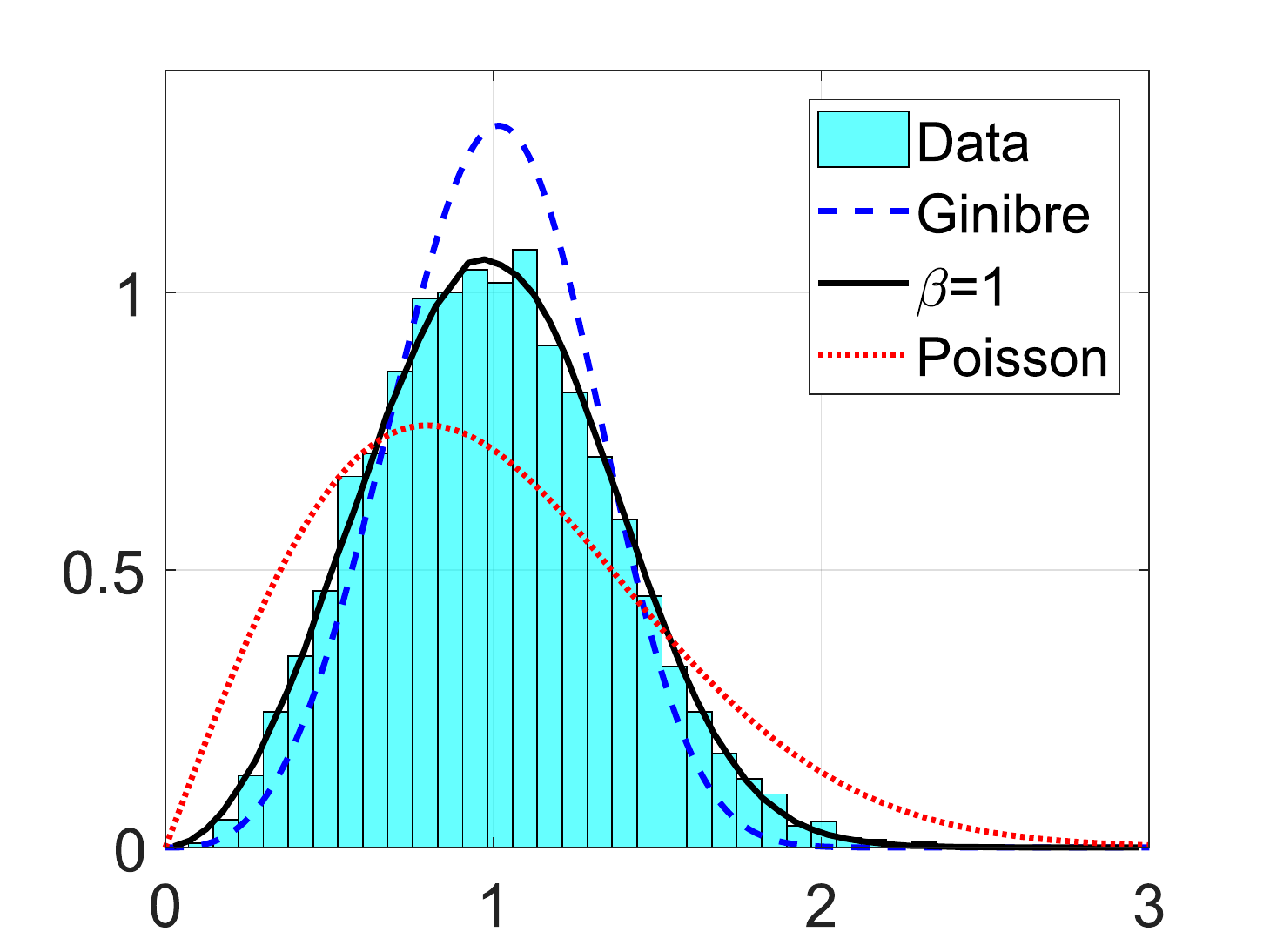}\\
		\hspace{-0.39\linewidth}	\textbf{(a)}\hspace{0.47\linewidth}	\textbf{(b)}\hfill\\
		\includegraphics[width=0.49\linewidth,angle=0]{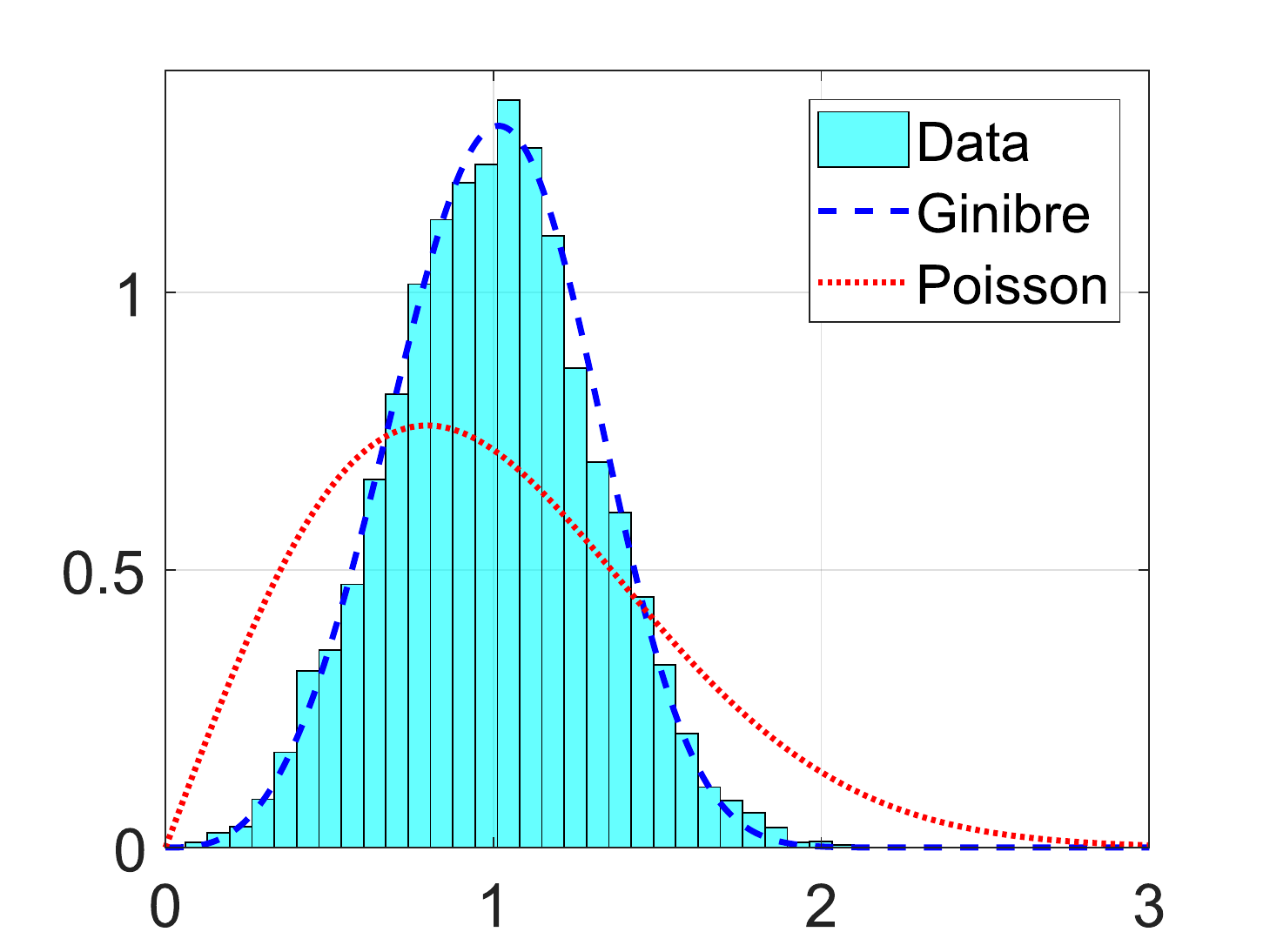}
		\includegraphics[width=0.49\linewidth,angle=0]{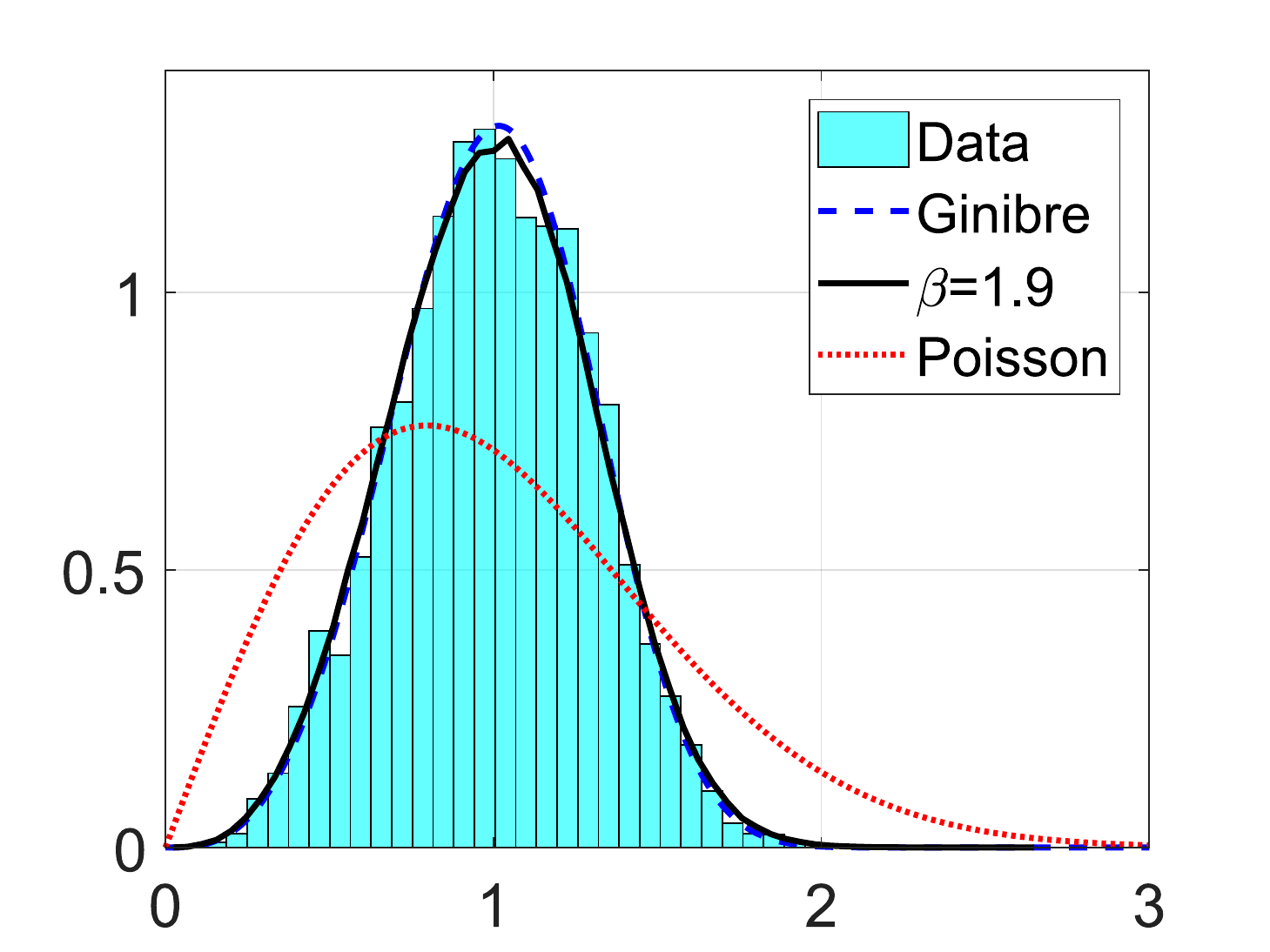}\\
		\hspace{-0.39\linewidth}	\textbf{(c)}\hspace{0.47\linewidth}	\textbf{(d)}\hfill
		\caption{{Comparison of
		the level spacing distributions for various Liouville operators~\eqref{Liouvillian},
			 	 the analytical spacing distributions~\eqref{Poisson-2d} (Poisson $\beta=0$, dotted) and ~\eqref{GinibreSpacing} (Ginibre $\beta=2$, dashed) as well as fits to general Coulomb gas~\eqref{2DCoulomb} simulations (Figures (b) and (d), solid).} Unfolding ~\eqref{unfold} is used with {the smearing parameter $\sigma=4.5\,\bar{s}$, see~\eqref{unfold}}, where the mean spacing varies from $\bar{s}=0.0036$ to $0.0045$ for the data sets (a) - (d). 
	The first moment of all spacings is normalised to unity.
%LiouvSpcDeph1
%0.0036
%AA
%0.0042
%A
%0.0045
%C
%0.0045	
}
		\label{Fig:Liouville}
	\end{figure}
We have generated four realisations of the Liouville operator~\eqref{Liouvillian} where for all four cases $N=10$ and $M=7$ and we set the scale to $J=1$. Thus, we have had in total $77520$ eigenvalues per case to analyse.
\begin{itemize}
\item[(a)]	The boundary driven XX-chain ($\Delta\!=\!0$) with bulk dephasing. The 
parameters are chosen as $J'=0$,
$\gamma_{\rm L}^+=0.5,\ \gamma_{\rm L}^-=1.2,\ \gamma_{\rm R}^+=\gamma=1,\ \gamma_{\rm R}^-= 0.8$. The model is equivalent to the Fermi--Hubbard chain with imaginary interaction $U = i\gamma$ with off-diagonal boundaries, see~\cite{MEP2016}, which is known to be Bethe ansatz integrable. According to the 
GHS-conjecture we expect Poisson statistics of the Liouvillian spectrum, see Fig.~\ref{Fig:Liouville}(a).
\item[(b)]	The isotropic Heisenberg XXX-chain ($\Delta=1$) with pure-source/pure-sink driving. The parameters are $J'=0,\ \gamma_{\rm L}^+=0.6,\ \gamma_{\rm R}^-=1.4,\ \gamma_{\rm L}^-=\gamma_{\rm R}^+=\gamma=0$ in this regime. The steady state (zero-mode) of this problem is known to be exactly-solvable~\cite{Prozen2011}, however the full Liouvillian spectrum shows non-integrable behaviour, see Fig.~\ref{Fig:Liouville}(b).
\item[(c)]	The XXX-chain  ($\Delta=1$) with arbitrary polarising boundary driving. Here, we chose the parameters $J' =0,\ \gamma_{\rm L}^+=0.5,\ \gamma_{\rm L}^-=0.3,\ \gamma_{\rm R}^+=0.3,\ \gamma_{\rm R}^-=0.9,\ \gamma=0$. The bulk Hamiltonian of this model is well-known to be integrable via Bethe ansatz, but with the boundary driving not even the steady state seems to be exactly solvable. The spectrum in Fig.~\ref{Fig:Liouville}(c) confirms that its dynamics is fully chaotic, according to the GHS-conjecture.
\item[(d)]	The XXZ-chain with nearest neighbour and next-to-nearest neighbour interactions. We have chosen $J' =1,\ \Delta=0.5,\ \Delta'=1.5$
with the same dephasing parameters as in (c). This time, even the bulk Hamiltonian is non-integrable (quantum chaotic) so that we expect Ginibre statistics 
following the GHS-conjecture, which is confirmed in Fig.~\ref{Fig:Liouville}(d).
\end{itemize}
\begin{table}[b!]
	\begin{tabular}{c||c|c|c}
		System & Poisson & fitted Coulomb  $\beta$& Ginibre \\
		\hline\hline
		(a) & 0.015& -- & 0.15\\
		\hline
		(b) & 0.10 & 0.0092 ($\beta=1$)& 0.058\\
		\hline
		(c) & 0.15 & --		& 0.012\\
		\hline
		(d) & 0.16 & 0.0094 ($\beta=1.9$)& 0.012
	\end{tabular}
	\caption{The Kolmogorov distance between the empirical data shown in Fig.~\ref{Fig:Liouville}, the Poisson distribution ~\eqref{Poisson-2d}, fitted value for $\beta$ (specified in the inset) of the Coulomb gas and the Ginibre spacing distribution~\eqref{GinibreSpacing}.}
	\label{Tab:Kolmogorov}
\end{table}
{All four data sets are depicted in Fig.~\ref{Fig:Liouville}, illustrating the integrable Fig.~\ref{Fig:Liouville}(a), intermediate  Fig.~\ref{Fig:Liouville}(b) and apparently fully chaotic  cases Figs.~\ref{Fig:Liouville}(c)-(d). 
%%%%
Note that the intermediate case (b) flows closer (and is expected to converge) to fully chaotic statistics by increasing the dimension $\kappa$.
%%%%
We compare with the 2D Poisson distribution~\eqref{Poisson-2d}, the distribution of the numerically generated Coulomb gas \eqref{2DCoulomb} with best fit for $\beta$, 
and the level spacing distribution~\eqref{GinibreSpacing} of the complex Ginibre ensemble. 
The Kolmogorov--Smirnov distances
~\cite{Kolmogorov} between the empirical distributions of the spectrum of ${L}$, 
and each of these curves (after fitting $\beta$) are listed in Table \ref{Tab:Kolmogorov}. 
The spacings for the Coulomb gas are obtained by generating points with the distribution \eqref{2DCoulomb} by using the Metropolis algorithm,
following \cite{Chafai}, and then determining the spacing numerically.  
Fig.~\ref{Fig:Liouville} confirms our expectations of an extended GHS-conjecture~\cite{GHS1988,GH1989} for dissipative open quantum systems to hold, even without classically chaotic correspondents}.\\

\paragraph{Unfolding of Complex Spectra.}

In order to compare the spectrum of ${L}$ with the spectral statistics of the 2D Coulomb gas \eqref{Poisson-2d}--\eqref{2DCoulomb} we need to unfold the spectrum. This means that we have to separate the fluctuations (fl), that are supposedly universal, from the global, averaged (av) spectral density which is system specific: 
\begin{equation}
\label{rhosplit}
\rho(x,y)={\sum}_{i=1}^N\delta^{(2)}(z-z_i)=\rho_{\rm av}(x,y)+ \rho_{\rm fl}(x,y)\ ,
\end{equation}
where $z=x+iy$. For real spectra unfolding is achieved by introducing the cumulative spectral function and fitting the smooth part $\eta(x)=\int_{-\infty}^{x}\rho_{\rm av}(t) dt$~\cite{GMGW}. For complex spectra this is more involved. 
Following \cite{Wettig}, unfolding is a map 
\begin{equation}
\label{map}
z\to z'=x'+iy'=u(x,y)+iv(x,y)
\end{equation}
to be found,
that satisfies certain conditions. First, after unfolding the density has to be unity (or constant), $\rho_{\rm av}(x',y')=1$, or in other words the Jacobian of the transformation \eqref{map} has to cancel the density before unfolding, $dx'dy'=\rho_{\rm av}(x,y)dxdy$. This is certainly not unique, and we believe that, second, local isotropy has to be achieved, e.g. using conformal maps \cite{Wettig}. 
{Following the symmetry of their data the authors \cite{Wettig} proposed to unfold in strips parallel to the $x$-axis, in choosing $y'=y$ and thus
$x'=\int_{-\infty}^{x}\rho_{\rm av}(t,y) dt=u(x,y) .$
{Apparently 
for more general data sets
this choice is not ideal, e.g. for products of $M$ Ginibre matrices where the density at the origin is singular \cite{Burda}.
}Its local statistics  is known to still follow the complex Ginibre ensemble \cite{ABu}, making proper unfolding crucial.

In fact we found a much simpler method following \eqref{rhosplit}, by approximating 
$\rho_{\rm av}(x,y)$ by a sum of Gaussian distributions around each eigenvalue $z_j$,
\begin{equation}
\label{unfold}
\rho_{\rm av}(x,y)\approx \frac{1}{2\pi\sigma^2 N}{\sum}_{j=1}^N\exp\biggl[\frac{-1}{2\sigma^2}|z-z_j|^2
\biggl].
\end{equation}
The measured spacing at a point $z_0$ is then simply multiplied by $\sqrt{\rho_{\rm av}(x_0,y_0)}$.
Testing this on spectra of products of random matrices, the choice $\sigma=4.5\,\bar{s}$ in terms of the global mean spacing $\bar{s}$ leads to very good results, see \cite{Supp}. This method is applied to our data sets (a) - (d) in Fig. \ref{Fig:Liouville}.
}\\

\paragraph{Random Matrix Universality.}

The question raised by the conjecture of GHS was why the fully chaotic case should be compared with the predictions of the complex Ginibre ensemble. They showed~\cite{GHS1988} that due to Hermiticity constraints generic dissipative open quantum systems lead to a spectrum of real and complex conjugate {eigenvalue} pairs. Thus one would expect the real or quaternion Ginibre ensemble (GinOE or GinSE) sharing this property to apply,  and not the GinUE. However, they found an agreement of their data from periodically kicked tops with damping with the GinUE - results for the GinOE or GinSE were not available at the time. 

While the results for the GinSE became available soon after \cite{Mehta}, including the spacing distribution at the origin (which is different from the GinUE \eqref{GinibreSpacing}), the GinOE was independently solved much later
by three groups \cite{HJS,Forrester08,BorodinSinclair}. They are given by so-called Pfaffian point processes, with matrix valued kernels as the main building block. 

Once all density correlation functions are known all spectral information is given, including the spacing. While close to the real line all three ensembles differ, it was shown that at the edge of the spectrum the GinSE \cite{Rider} and GinOE \cite{BorodinSinclair} agree with the GinUE \cite{ForresterHonner}.
It is therefore natural to ask if this agreement continues to hold in the bulk  or not.
For the GinOE this was answered affirmatively  in \cite{BorodinSinclair}, and in the supplement of the present work \cite{Supp} which includes
Refs. \cite{NIST,EK} we show that this also holds for the GinSE. 
Below we give a heuristic argument (see also \cite{Haake}), why 
all three symmetry classes yield the same spacing distribution in the bulk 
and it is  thus universal.

The joint probability density function (jpdf) of eigenvalues for all three Ginibre ensembles read~\cite{LS1991,Ginibre}
\begin{eqnarray}
\nonumber
&&\mathcal{P}_{\rm GinOE}^{(k)}(z)\! \propto \!
|\Delta_M(z)|^2\Delta_k(x)\!\!\prod_{i,j=1}^M\!(z_i-z_j^*)\!\prod_{i=1}^k\prod_{j=1}^M|z_j-x_i|^2\\
&&\times\prod_{l=1}^{k}  e^{-\frac12x_l^2}\prod_{j=1}^{M} {\rm sign}({\rm Im}(z_j))\erfc\big(\sqrt{2}{\rm Im}(z_j)\big)e^{-\frac12(z_j^2+z_j^{*\,2})}\!,
\nonumber\\
&&\mathcal{P}_{\rm GinUE}(z)\! \propto |\Delta_N(z)|^2 {\prod}_{j=1}^{N} e^{-|z_j|^2},
\label{jpdfs}\\
&&\mathcal{P}_{\rm GinSE}(z)\! \propto 
|\Delta_M(z)|^2 
\prod_{i>j}^M|z_i-z_j^*|^2\!
\prod_{j=1}^{M}|z_j-z_j^*|^2 e^{-|z_j|^2}\!.
\nonumber
\end{eqnarray}
Here, $\Delta_N(a)=\prod_{j>k}^N(a_j-a_k)$ denotes the Vandermonde determinant and  
$\erfc$ the complementary error function. The $N=k+2M$ eigenvalues in the GinOE are ordered to yield a positive density, and $k$ counts the number of real eigenvalues, see e.g. \cite{HJS,Forrester08, BorodinSinclair} for details, and for the GinSE $N=2M$.

\begin{center}
	\begin{figure}[t]
		\centering
		\includegraphics[width=0.8\linewidth,angle=0]{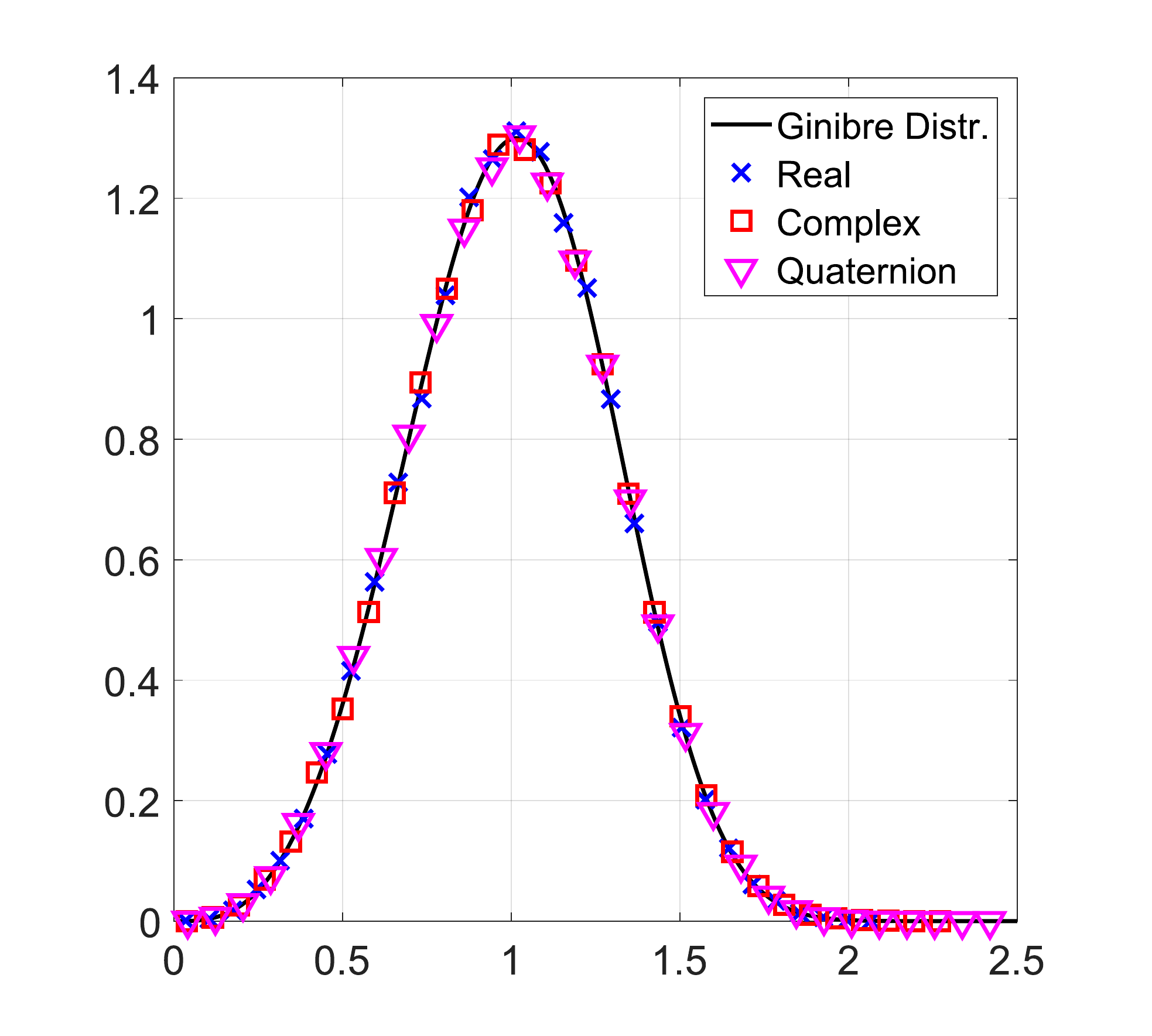}
		\caption{Comparison of the	
		spacing distribution~\eqref{GinibreSpacing} 
		with 		normalised first moment  
		and those of
				the GinOE (blue crosses), GinUE (red squares), and GinSE (purple triangles)
		in the bulk of the spectrum. 
%%%%%%%%%
For the latter we use the standard $2N$-\newline dimensional representation of an $N$-dimensional quaternionic matrix, making the complex eigenvalues unique, cf. \cite{Ginibre}.
An ensemble of $1000$ $500\times 500$ matrices has been generated in a Monte Carlo simulation.
	 Here, the 
	 unfolding is trivial due to a uniform density
	 of all three ensembles.
		}
		\label{Fig:GinibreCompare}
	\end{figure}
\end{center}

For large-$N$ there are only $k\propto\sqrt{N}$ real eigenvalues $x_l$ on average \cite{Efetov}, and thus we consider $M\sim N/2$. Raising the Vandermonde to the exponent leads to the Coulomb gas picture \eqref{2DCoulomb} at $\beta=2$ for the GinUE.
Notice that the other 2 ensembles are {\it not} proportional to $|\Delta_N(z)|^\beta$ for $\beta=1,4$.
The limiting spectral density is constant on a disc of radius {$\mathcal{O}(\sqrt{N})$ for all three} Ginibre ensembles, and also for Coulomb gases \eqref{2DCoulomb} {for all $\beta>0$,} see e.g. \cite{Serfaty} for a review.
The local bulk statistics is defined by zooming into the vicinity
%%%%%%%%%%%%%
of radius $R=\mathcal{O}(1)$ of a few mean level spacings
%%%%%%%%%%%%%
around a bulk eigenvalue $z_0$, chosen far away from the real axis and the edge of the support. 
Close to $z_0$, complex conjugate and real eigenvalues are of the order $\mathcal{O}(\sqrt{N})$ away from $z_0$ and thus do not contribute to the local spectral statistics.
Hence all jpdfs \eqref{jpdfs} become locally proportional to
\begin{equation}
\sim
\prod_{j: |z_j-z_0|<R}
|z_j-z_0|^2
\end{equation}
for large $N$.
Thus all three ensembles coincide locally, and share the GinUE spacing distribution \eqref{GinibreSpacing}.
In Fig.~\ref{Fig:GinibreCompare}, we illustrate this argument with Monte-Carlo simulations of all three Ginibre ensembles in the bulk, finding perfect agreement.
{Very recently numerical evidence has been given for four further symmetry classes to follow the spacing \eqref{GinibreSpacing} of the GinUE \cite{H2}. While the authors identified 2 ensembles where the spacing differs, it remains to be seen  how many classes emerge in the bulk from the complete list of non-Hermitian ensembles \cite{LeClair,Magnea,H0}.
}\\[-6ex]
\paragraph{Conclusions}
We have studied universal spectral properties of dissipative open quantum systems. Their corresponding Liouville operator $L$ generically exhibits complex eigenvalue statistics. In our example we numerically diagonalised boundary driven quantum spin-chains of the XXZ type, with nearest and next-to-nearest neighbour interactions with different sets of couplings. Depending on these parameters, it is known that the system undergoes  a transition from integrable to chaotic behaviour. 
The spacing distribution in radial distance between the complex eigenvalues of $L$ has shown to be an efficient measure to observe this transition. Generalising the conjecture of Grobe, Haake and Sommers for the extreme cases, we have shown that the intermediate statistics is very well described by a two-dimensional Coulomb gas with harmonic potential, by fitting to an inverse temperature $\beta\in[0,2]$. Furthermore, we have generalised the universality argument of these authors from a cubic repulsion for small spacing in the chaotic case $\beta=2$ to hold for the full distribution in all three Ginibre ensembles. Here, we contributed analytically to the quaternion case, and illustrated this by numerical evidence. 

Several open questions deserve further studies. While for quantum systems with real eigenvalues the emergence of random matrix statistics in the chaotic regime is well understood, using a semi-classical expansion, such an approach is not developed here. 
Further examples for physical systems with complex eigenvalues 
should be studied throughout the transition region from integrable to chaotic behaviour, to see if the description by a 2D Coulomb gas is indeed universal.

\acknowledgements
\paragraph{Acknowledgements.} This work was partly funded by the Wallenberg foundation (GA), the German Science Foundation DFG within 
CRC1283 "Taming uncertainty and profiting from randomness and low regularity in analysis, stochastics and their applications" {(GA, MK)} and within IRTG2235 "Searching for the regular in the irregular: Analysis of singular and random systems" (AM) {, by European Research Council under the Advanced Grant No. 694544 -- OMNES (TP), and by 
the Slovenian Research Agency (ARRS) under the Programme P1-0402 (TP).
Support from the Simons Center for Geometry and Physics, Stony Brook University,  
is gratefully acknowledged 
where part of this work was completed
(GA, MK), as well as fruitful discussions with Maurice Duits (GA).}


\begin{thebibliography}{99}
\bibitem{betalim}
The rescaling is made for the limit $\beta\to0$ to exist, leading to non-interacting particles confined by a Gaussian potential.
\end{thebibliography}
\end{document}

\title{
{\it  Supplementary Material to:}
Universal Signature from Integrability to Chaos in Dissipative Open Quantum Systems}

\author{Gernot Akemann}\email{akemann@physik.uni-bielefeld.de}\affiliation{Faculty of Physics, Bielefeld University, Postfach 100131, 33501 Bielefeld, Germany,\\
Department of Mathematics, Royal Institute of Technology (KTH), 
Brinellv\"agen 8, 114 28 Stockholm, Sweden}
\author{Mario Kieburg}\email{m.kieburg@unimelb.edu.au}\affiliation{School of Mathematics and Statistics, University of Melbourne, 813 Swanston Street, Parkville, Melbourne VIC 3010, Australia}
\author{Adam Mielke}\email{amielke@math.uni-bielefeld.de}\affiliation{Faculty of Physics, Bielefeld University, Postfach 100131, 33501 Biele\-feld, Germany}\author{Toma\v{z} Prosen}\email{tomaz.prosen@fmf.uni-lj.si}\affiliation{Physics Department, Faculty of Mathematics and Physics, University of Ljubljana, Ljubljana 1000, Slovenia}

\date{\today}
\maketitle

\begin{appendix}
%%%%%%%%%%%%%%%%%%%%%%%%%%%%%%%%%%%%%%%%%%%%%%%%%%%%%%%%%%
\section{Unfolding and Spacings of the Coulomb Gas}

In this section we give some more details about our unfolding procedure and provide examples that illustrate our method. We then give examples for the numerically generated spacing distribution for the two-dimensional (2D) Coulomb gas at various values of $\beta$.

The goal of unfolding is to separate the fluctuations (fl) of the density from the global, averaged (av) spectrum 
\begin{equation}
\label{rhosplit}
\rho(x,y)=\sum_{i=1}^N\delta^{(2)}(z-z_i)=\rho_{\rm av}(x,y)+ \rho_{\rm fl}(x,y)\ ,
\end{equation}
where $z=x+iy$. In the new coordinates 
$z\to z'=x'+iy'$ the mean density should become uniform $dx'dy'=\rho_{\rm av}(x,y)dx\,dy$, and the fluctuations around it are to be compared with the universal predictions from random matrices. 

Our procedure to determine the average density is very simple.  
The Dirac delta at each eigenvalue $z_j$ is replaced by a broadened Gaussian distribution 
\begin{equation}
\label{unfold}
\rho_{\rm av}(x,y)\approx \frac{1}{2\pi\sigma^2 N}\sum_{j=1}^N\exp\left[\frac{-1}{2\sigma^2}|z-z_j|^2
\right].
\end{equation}
We choose the variance  $\sigma$ to be larger than the global mean spacing $\bar{s}$ between the eigenvalues, in order to obtain a smooth density. 
The unfolded spacing is then obtained for every bulk eigenvalue $z_j$ by finding its nearest neighbour, and then multiply their distance  by 
$\sqrt{\rho_{\rm av}(x_j,y_j)}$, as
the inverse local mean density measures the local area. 
The resulting empirical spacing distribution is then normalised by rescaling the first moment to unity. This is compared to the equally normalised spacing distribution  of the complex Ginibre ensemble (GinUE), 
\begin{equation}
p(s)=d\, p_{{\rm GinUE}}(ds), \quad 
d=\int_0^\infty ds\,s\, p_{{\rm GinUE}}(s)\approx 1.14, \quad 
p_{{\rm GinUE}}(s)=
\sum_{j=1}^{\infty}\frac{2s^{2j+1}e^{-s^2}}{\Gamma(j+1,s^2)} 
\prod_{j=1}^{\infty}\frac{\Gamma(j+1,s^2)}{\Gamma(j+1)}\ .
\label{1stmom}
\end{equation}
	\begin{figure}[b!]
		\centering
		\includegraphics[width=0.39\linewidth,angle=0]{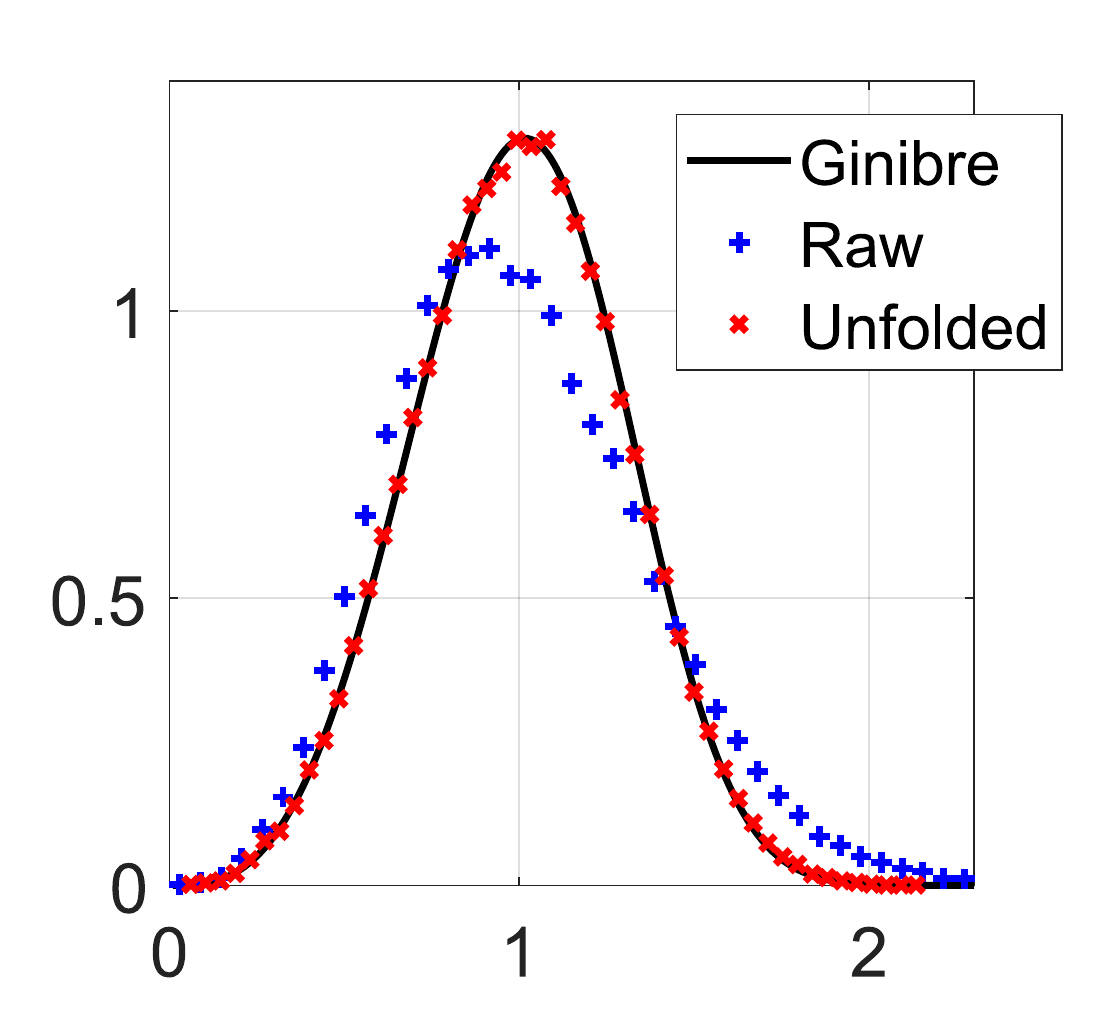}
		\includegraphics[width=0.39\linewidth,angle=0]{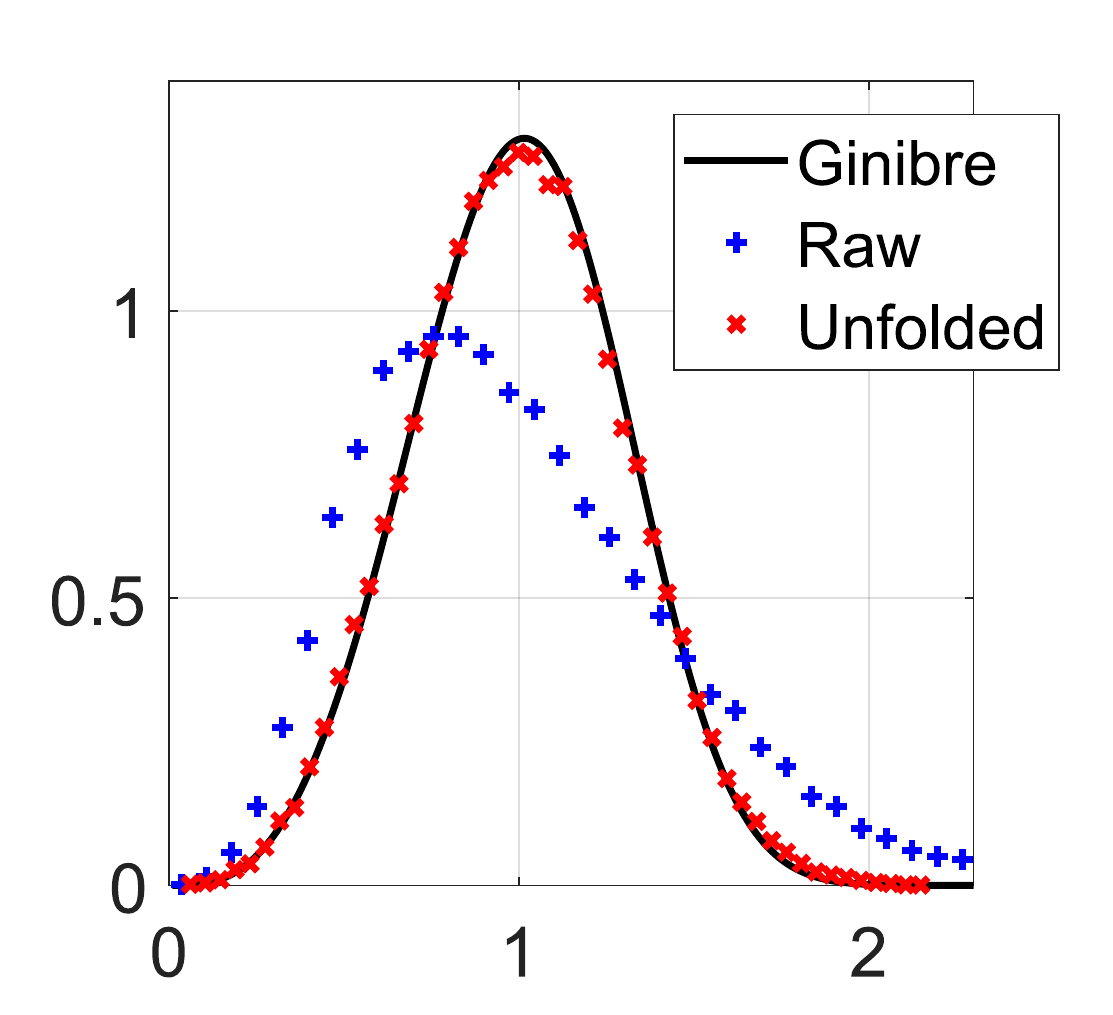}\\
		\caption{Comparison of the spacing distribution in the bulk of the spectrum for the product of $M=2$ (left) and $M=4$ (right) complex Ginibre matrices of sizes $N=200$, showing data from 1000 ensembles. Blue plus symbols indicate the data before and red crosses after unfolding, the full line gives the rescaled Ginibre distribution \eqref{1stmom}.}
		\label{Fig:Product}
	\end{figure}
We have tested the above procedure for the spacing distribution of the eigenvalues of the product of $M=2,3,4,5$ matrices from the GinUE, because the result is known analytically. They provide a good testing ground as their
average mean density is not constant \cite{Burda}
\begin{equation}
\rho_{\rm av}(x,y)=
 \frac{1}{M\pi }|z|^{2(1-M)/M}\ \Theta(1-|z|)\ ,
 \label{products}
\end{equation}
making unfolding necessary. Furthermore, it is known that the local statistics in the bulk of the spectrum agrees with that of the GinUE \cite{ABu}, and hence with its spacing distribution \eqref{1stmom}. 
Because of this we have used these ensembles to find a good choice for $\sigma$ in \eqref{unfold}. The choice $\sigma=4.5\,\bar{s}$ leads to very good results, as shown in Fig. \ref{Fig:Product}. 

For treating the data from the Liouville operator we proceed in the same way, by first computing the global average spacing $\bar{s}$, then determining the average density according to \eqref{unfold}, and finally normalising the first moment to unity.
\\

	\begin{figure}[t!]
		\centering
		\includegraphics[width=0.39\linewidth,angle=0]{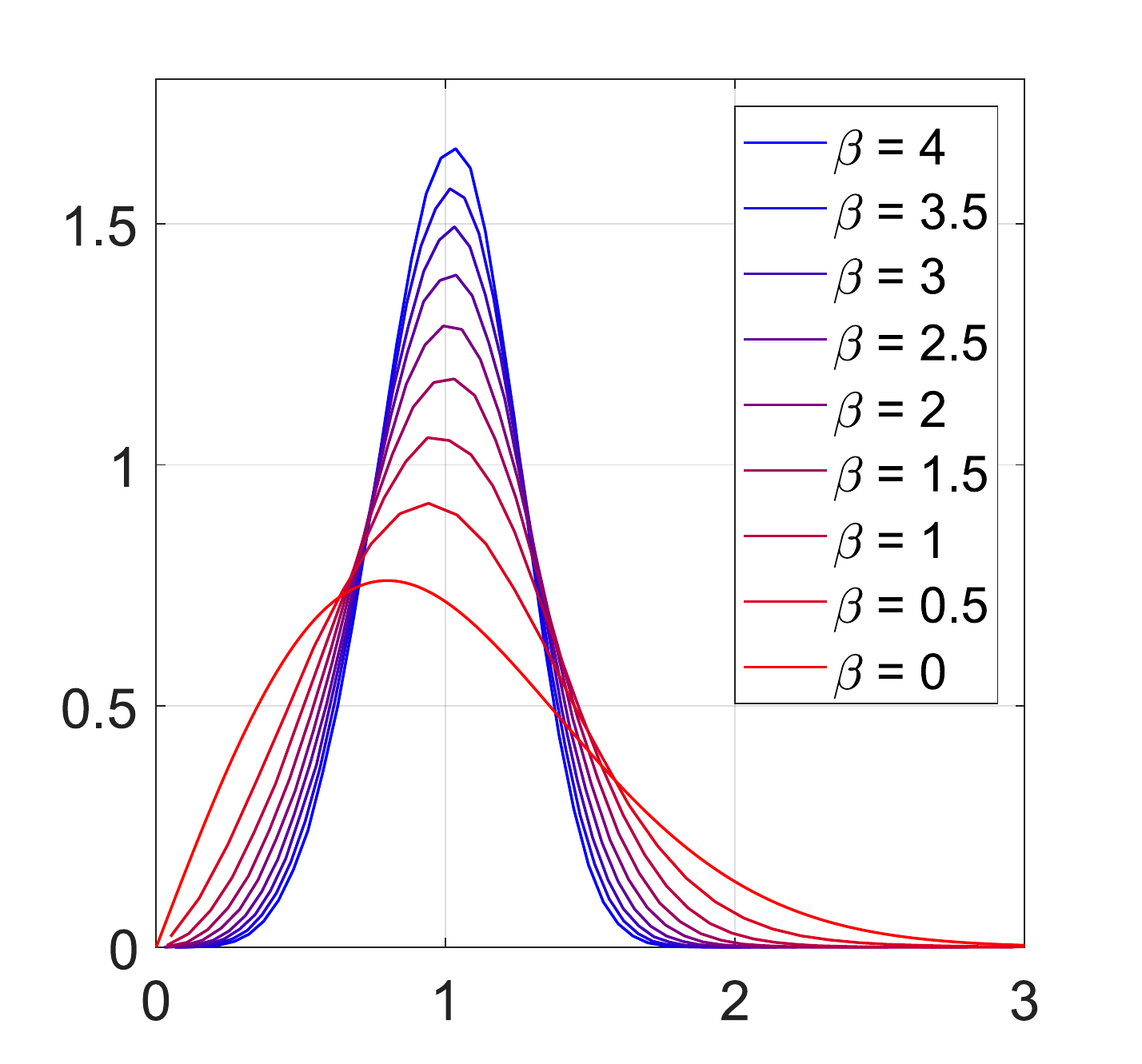}
		\caption{Numerically generated spacing distribution $p(s)$ for a 2D Coulomb gas for various values of $\beta=0.5,\ldots,4$. Also shown is the Poisson distribution \eqref{Poisson-2d} in 2D corresponding to $\beta=0$, with the weakest (linear) repulsion for small $s$.}
		\label{Fig:Coulomb}
	\end{figure}
The second purpose of this section is to give examples for the spacing distribution of the 2D Coulomb gas. Its unnormalised joint distribution of charges in the plane reads
\begin{equation}
\mathcal{P}_{\rm Coulomb}(\{z\}) =\prod_{j>k}^N|z_j-z_k|^\beta\prod_{j=1}^{N} e^{-|z_j|^2},\quad \beta>0\ .
\label{jpdf}
\end{equation}
The points distributed according to this distribution are generated following \cite{Chafai}, using a Metropolis algorithm. 
Here, unfolding is a trivial rescaling because the corresponding Coulomb gas is known to follow the circular law, \eqref{products} at $M=1$, for arbitrary $\beta>0$, cf. \cite{Serfaty} for a review.
We have generated a library of spacings in steps of $0.1$ in $\beta$, that was used to fit the data from the Liouville operator. In Fig. \ref{Fig:Coulomb} we give some examples for the so obtained spacings, where we have normalised the first moment to unity for each curve. Only for $\beta=2$ the analytical result is known  \eqref{1stmom}, as well as for $\beta=0$ which yields the Poisson distribution in 2D, 
\begin{equation}\label{Poisson-2d}
p_{\rm P}^{\rm (2D)}(s)= \frac{\pi}{2} s e^{-\pi s^2/4} .
\end{equation}
Note that the first moment of the Poisson distribution is already unity. It is also displayed in Fig. \ref{Fig:Coulomb} for comparison.

%%%%%%%%%%%%%%%%%%%%%%%%%%%%%%%%%%%
\section{Bulk Universality of the Symplectic Ginibre Ensemble}

\subsection{Complex Ginibre Ensemble}
In order to state the universality of the symplectic Ginibre ensemble (GinSE) in the bulk of the spectrum, we recall what is known for the complex Ginibre ensemble, to which we would like to show equivalence. We follow here \cite{Ginibre,Mehta}.
The joint probability density function (jpdf) of the GinUE equals \eqref{jpdf} at $\beta=2$, 
 $\mathcal{P}_{\rm GinUE}(\{z\})=\mathcal{P}_{\rm Coulomb}(\{z\})|_{\beta=2}$,
with normalising partition function $Z_N=N!\ \pi^N\prod_{j=0}^{N-1} j!$. 
The $k$-point density correlation functions defined as 
\begin{equation}
\label{Rk}
R_{k,N}^{\rm GinUE}(z_1,\ldots,z_k)\equiv\frac{1}{Z_N}  \frac{N!}{(N-k)!}\int_{\mathbb{C}^{N-k}}d^2z_{k+1}\cdots d^2z_N \mathcal{P}_{\rm GinUE}(\{z\}) =\det_{1\leq i,j\leq k}[K_N(z_i,z_j^*)]
\end{equation}
form a determinantal point process (DPP) at finite $N$. Its kernel of orthogonal polynomials is given by 
\begin{equation}
\label{kernel}
K_N(z,u^*)=\frac{1}{\pi} e^{-\frac{|z|^2}{2}-\frac{|u|^2}{2}}\sum_{j=0}^{N-1}\frac{(zu^*)^j}{j!}= \frac{1}{\pi} e^{-\frac{|z|^2}{2}-\frac{|u|^2}{2}} e_{N-1}(zu^*)\ ,
\end{equation}
and determines all spectral correlation functions for finite-$N$.
It contains $e_{N-1}(x)=\sum_{j=0}^{N-1}x^j/j!$, the exponential polynomial  of degree $N-1$.
The large-$N$ limit at the origin, with $z,u=\mathcal{O}(1)$, is trivial and given by 
\begin{equation}
K_{\rm Gin}(z,u^*)=\lim_{N\to\infty}K_N(z,u^*) = \frac{1}{\pi} \exp\left[-\frac{|z|^2}{2}-\frac{|u|^2}{2}+zu^*\right],
\label{DefKGin}
\end{equation}
the limiting Ginibre kernel.
It determines all limiting $k$-point correlation functions at the origin.
The limit in the bulk of the spectrum is based on the asymptotic of the exponential polynomial, expressed in terms of the incomplete Gamma-function. For the latter the following uniform asymptotic holds \cite{NIST} 
\begin{equation}
\lim_{N\to\infty}\left(\frac{\Gamma(N,Nx)}{\Gamma(N)}=
e^{-Nx}e_N(Nx)\right)=
\left\{
\begin{array}{cc}
0, & x>1\ ,\\
1,& 0\leq x<1\ .
\end{array}
\right.
\label{e_n-lim}
\end{equation}
This implies the limiting circular law 
$R_{1,N}^{\rm GinUE}(z)= \frac1\pi e^{-|z|^2} e_{N-1}(|z|^2)\sim\frac{1}{\pi}\Theta(1-|z|^2/N)$,
that the macroscopic (or global) spectral density is constant on a disc of radius $\sqrt{N}$.
For the local bulk limit we zoom into the vicinity of the bulk point $\sqrt{N}z_0$, with 
$|z_0|<1$ to stay away from the edge. From \eqref{e_n-lim} we can thus  replace the exponential polynomial by an exponential, to obtain the following kernel at large-$N$
\begin{equation}
K_N\left(\sqrt{N}z_0+\xi,\sqrt{N}z_0^*+\eta^*\right)
\sim
\frac{f_N(\xi)}{f_N(\eta)} K_{\rm Gin}(\xi,\eta^*)\ ,
\quad \mbox{with}\ \ 
 f_N(\xi)=\exp\left[\frac12\sqrt{N}(z_0^*\xi-z_0\xi^*)\right]\ ,
 \label{f-def}
\end{equation}
and $\xi,\eta=\mathcal{O}(1)$. 
Noting that the $k$-point correlation functions remain unchanged when replacing the kernel by an equivalent kernel, $K_N(z,u)\to\frac{g(z)}{g(u)}K_N(z,u)$ for an arbitrary $g(z)\neq0$ due to the invariance of the determinant, 
the following bulk limit holds for all $k$-point correlation functions
\begin{equation}
\lim_{N\to\infty}R_{k,N}^{\rm GinUE}\left(\sqrt{N}z_0+\xi_1,\ldots,\sqrt{N}z_0+\xi_k\right)=\det_{i,j=1,\ldots,k}\left[K_{\rm Gin}(\xi_i,\xi_j^*)\right]\ .
\label{Rk-Gin-b}
\end{equation}
Because all $k$-point correlation functions at the origin 
and in the bulk agree, also all other spectral correlation functions will agree, including the limiting spacing distribution \eqref{1stmom}.

%%%%%%%%%%%%%%%%%%%%%%%%%%%%%%%%%%%%%%%%%%%%%%%%%%%%%
\subsection{Symplectic Ginibre Ensemble}

Following \cite{Mehta,EK} the GinSE is given by the jpdf of $N$ complex eigenvalues and normalisation constant
\begin{equation}
\mathcal{P}_{\rm GinSE}(\{z\})=
\prod_{i>j}^N|z_i-z_j|^2|z_i-z_j^*|^2
\prod_{l=1}^{N}|z_l-z_l^*|^2\ e^{-|z_l|^2}\ ,
\quad Z_N=N!(2\pi)^N\prod_{j=1}^{N}(2j-1)!\ ,
\label{jpdfSE}
\end{equation}
where $N=2M$. It
has the following $k$-point correlation functions defined as in \eqref{Rk}
\begin{equation}
R_{k,N}^{\rm GinSE}(z_1,\ldots,z_k)=
\prod_{l=1}^k(z_l^*-z_l)
\mbox{Pf}_{i,j=1,\ldots,k}\left[\mathcal{K}_N(z_i,z_j)\right]\ ,
\label{Rk-qGin}
\end{equation}
forming a Pfaffian point process with the matrix valued $2\times2$ 
kernel
\begin{equation}
\mathcal{K}_N(z,w)=e^{-(|z|^2+|w|^2)/2}
\left(
\begin{array}{cc}
\kappa_N(z,w) & \kappa_N(z,w^*)\\
\kappa_N(z^*,w)     & \kappa_N(z^*,w^*)\\
\end{array}
\right).
\label{K2x2}
\end{equation}
The latter contains a single antisymmetric pre-kernel at different arguments,
\begin{equation}
\kappa_N(z,w)  =\frac{1}{\pi}\sum_{k=0}^{N-1}
\left(\frac{z^{2k+1}}{(2k+1)!!}\sum_{j=0}^k\frac{w^{2j}}{(2j)!!}
-\frac{w^{2k+1}}{(2k+1)!!}\sum_{j=0}^k\frac{z^{2j}}{(2j)!!} 
\right).
\label{kappa-n}
\end{equation}
Clearly all density correlation functions \eqref{Rk-qGin} vanish along the real axis. Notice that we have not pulled the pre-factor in \eqref{Rk-qGin} inside the Pfaffian determinant, in order to avoid taking unnecessary square roots.

It is not difficult to show that the pre-kernel
satisfies the following system of differential equations at finite-$N$:
\begin{eqnarray}
\partial_z \kappa_N(z,w) -z\, \kappa_N(z,w) &=&\ \ \frac{1}{\pi} e_{2N-1}(zw) -
\frac{1}{\pi} \frac{z^{2N}}{(2N-1)!!}e_{N-1}(w^2/2) \ ,\nonumber\\
\partial_w \kappa_N(z,w) - w\,\kappa_N(z,w) &=& - \frac{1}{\pi} 
e_{2N-1}(zw) +
\frac{1}{\pi} \frac{w^{2N}}{(2N-1)!!}e_{N-1}(z^2/2) 
\ .
\label{PDEsN}
\end{eqnarray}
Following \eqref{e_n-lim}  the exponential polynomials on the right hand side (rhs) can be replaced by exponentials, as long as $|z|,|w|\leq C<\sqrt{2N}$, 
{
allowing them to diverge, $|z|,|w|\to\infty$ as $N\to\infty$.}
This is compatible with the limiting spectral density obeying the circular law on a disc of radius $\sqrt{2N}$ (compared to $\sqrt{N}$ for the GinUE). 

We will show that the second term on both rhs' in \eqref{PDEsN} are subleading for large-$N$. {For $z,w=\mathcal{O}(1)$ this is obvious, using Stirling's formula.} Defining 
\begin{equation}
\kappa_N(z,w)=e^{\frac{1}{2}(z^2+w^2)}\Lambda_N(z,w)\ ,
\end{equation}
the lhs of the first equation \eqref{PDEsN} becomes $e^{\frac{1}{2}(z^2+w^2)}\partial_z\Lambda_N(z,w)$, and likewise for the second. Putting the exponentials on the other side 
and integrating, we obtain for large-$N$
\begin{eqnarray}
\Lambda_N(z,w)\sim
\frac{1}{\pi}\int_w^z du\,e^{-\frac12(u-w)^2} -\frac{1}{\pi}\int_w^z du\frac{1}{\sqrt{2}}\left(\frac{u^2e}{2N}\right)^N e^{-\frac12 u^2}
=\frac{1}{\sqrt{2\pi}} \mbox{erf}[(z-w)/\sqrt{2}] -\mathcal{I}(z,w) \ .
\label{LaN}
\end{eqnarray}
Here, we have used 
$\Lambda_N(w,w)=0$ due to antisymmetry, applied Stirling's formula and introduced  the error function,
$\mbox{erf}(x)=\frac{2}{\sqrt{\pi}}\int_0^x e^{-t^2}dt$, which is antisymmetric, too. The integral $\mathcal{I}(z,w)$ defined in \eqref{LaN} can be estimated by the absolute value of the integrand, 
{parametrising $u=(1-t)w+tz=x+iy$ with $t\in[0,1]$, and applying the ML Lemma:
\begin{eqnarray}
|\mathcal{I}(z,w)|\leq \frac{1}{\pi\sqrt{2}}\int_0^1dt |z-w|\left(\frac{(x^2+y^2)e}{2N}\right)^N e^{-\frac12(x^2+y^2)}e^{y^2}
\leq\frac{1}{\pi\sqrt{2}}|z-w|\max_{0\leq t\leq1}\left\{e^{N(\log (q(t)^2)+1-q(t)^2)}\right\}e^{a} .\quad\quad
\label{Iint}
\end{eqnarray}
Here, we use that when $|z|,|w|$ diverge as $\sim\sqrt{2N}$, also $u$ diverges as $|u(t)|=\sqrt{2N}q(t)$, with $0<q(t)<1$. This leads to an estimate for the first part of the integrand, with a negative exponent. The remaining factor $e^{y^2}$ has been estimated 
defining $a=\max\{ [\im(z)]^2,[\im(w)]^2\}$.
It is compensated after multiplying back with $\exp[(z^2+w^2)/2]$ and the weight from \eqref{K2x2}:
\begin{equation}
\left|\exp\left[-\frac12(|z|^2+|w|^2-z^2-w^2)+a\right]\right|=
\left|\exp[-[\im(z)]^2-[\im(w)]^2+a]\right|\leq \exp[-b],
\end{equation}
for $b=\min\{[\im(z)]^2,[\im(w)]^2\}\to\infty$ equally diverging for $N\to\infty$. Thus this contribution to the pre-kernel vanishes exponentially for large $N$, even if $|z-w|$ may diverge as $\mathcal{O}(\sqrt{2N})$.
The analysis for the second equation in \eqref{PDEsN} follows from interchanging $z\leftrightarrow w$. 
In summary, for large-$N$ we obtain the following asymptotic on the domain $|z|,|w|\leq C<\sqrt{2N}$,
\begin{equation}
e^{-(|z|^2+|w|^2)/2}\kappa_N(z,w)\sim e^{-(|z|^2+|w|^2)/2}\kappa(z,w)\ ,
\end{equation}
where we defined 
\begin{equation}
\kappa(z,w)=\frac{1}{\sqrt{2\pi}} \exp[(z^2+w^2)/2]\ \mbox{erf}[(z-w)/\sqrt{2}] \ .
\label{kappa-sol}
\end{equation}
For $z,u=\mathcal{O}(1)$ this was derived in \cite{EK}, corresponding to the large-$N$ limit at the origin. 
}
There, the limiting correlation functions read
\begin{equation}
\lim_{N\to\infty}R_{k,N}^{\rm GinSE}(z_1,\ldots,z_k)=
\prod_{l=1}^k(z_l^*-z_l)
\mbox{Pf}_{i,j=1,\ldots,k}\left[
e^{-(|z_i|^2+|z_j|^2)/2}
\left(
\begin{array}{cc}
\kappa(z_i,z_j) & \kappa(z_i,z_j^*)\\
\kappa(z_i^*,z_j)     & \kappa(z_i^*,z_j^*)\\
\end{array}
\right)
\right]\ .
\label{Rk-origin}
\end{equation}
These are clearly different from the $k$-point correlations of the GinUE \eqref{Rk-Gin-b}, e.g. the local density of the GinSE at the origin is given by 
\begin{equation}
\label{R1-origin}
\lim_{N\to\infty}R_{1,N}^{\rm GinSE}(z)=
(z^*-z)  e^{-|z|^2}\kappa(z,z^*)
=\frac{1}{\sqrt{2\pi}}(z^*-z)  \exp[-|z|^2+(z^2+z^{*\,2})/2]\ \mbox{erf}[(z-z^*)/\sqrt{2}] \ .
\end{equation}

To prove the agreement of all $k$-point correlations with the GinUE in the bulk of the spectrum, away from the real axis, we will proceed as in the bulk limit of the GinUE in the previous sub-section, cf. \eqref{f-def}, by using the uniform limit to the solution \eqref{kappa-sol}.  
We choose a bulk point $\sqrt{2N}z_0$ of the spectrum, with $|z_0=x_0+iy_0|<1$, and $z_0-z_0^*=\mathcal{O}(1)$ to stay away from the real axis, and we choose arguments $\sqrt{2N}z_0+\xi$ and $\sqrt{2N}z_0+\eta$ of the pre-kernel being in the vicinity of $\sqrt{2N}z_0$. Due to the structure of the Pfaffian \eqref{Rk-qGin} we have to distinguish four different combinations of complex conjugations for the arguments of the pre-kernel, times the weight function. 

We first take the limit when none of the arguments is complex conjugated: 
\begin{eqnarray}
e^{-\frac12(|\sqrt{2N}z_0+\xi|^2+|\sqrt{2N}z_0+\eta|^2)}
\kappa_N(\sqrt{2N}z_0+\xi,\sqrt{2N}z_0+\eta)
&\sim& \frac{1}{\sqrt{2\pi}}
e^{-\frac12(|\sqrt{2N}z_0+\xi|^2+|\sqrt{2N}z_0+\eta|^2
-(\sqrt{2N}z_0+\xi)^2-(\sqrt{2N}z_0+\eta)^2
)}
\nonumber\\
&&\times\mbox{erf}[(\sqrt{2N}z_0+\xi-\sqrt{2N}z_0-\eta)/\sqrt{2}] \ .
\label{kappa-zw}
\end{eqnarray}
The $N$-dependence inside the error function cancels, with the argument of $\mbox{erf}[(\xi-\eta)/\sqrt{2}] $ being of order one.
Multiplying out we obtain for the exponential factors
\begin{eqnarray}
\exp\left[
-\frac12\left(4N(|z_0|^2-z_0^2) + \sqrt{2N}\left((z_0^*-2z_0)(\xi+\eta)+z_0(\xi^*+\eta^*)\right)
+|\xi|^2-\xi^2+|\eta|^2-\eta^2
\right)
\right].
\end{eqnarray}
Obviously the leading term in the exponent $-4N(y_0^2-ix_0y_0)$ will make the limit of this matrix element vanish,
\begin{equation}
\lim_{N\to\infty}e^{-\frac12(|\sqrt{2N}z_0+\xi|^2+|\sqrt{2N}z_0+\eta|^2)}
\kappa_N(\sqrt{2N}z_0+\xi,\sqrt{2N}z_0+\eta) =0\ .
\end{equation}
Despite the invariance of the Pfaffian determinant $\mbox{Pf}(B^T AB)=\det(B)\mbox{Pf}(A)$, allowing to go to an equivalent matrix kernel, such transformations cannot compensate the leading order term as it is $\xi$- and $\eta$-independent. The same argument applies to the pre-kernel with both arguments complex conjugated, as it is trivially obtained via complex conjugation:
\begin{equation}
\lim_{N\to\infty}e^{-\frac12(|\sqrt{2N}z_0+\xi|^2+|\sqrt{2N}z_0+\eta|^2)}
\kappa_N(\sqrt{2N}z_0^*+\xi^*,\sqrt{2N}z_0^*+\eta^*) =0\ .
\end{equation}
Let us come to the pre-kernel with one complex conjugated argument only: 
\begin{eqnarray}
e^{-\frac12(|\sqrt{2N}z_0+\xi|^2+|\sqrt{2N}z_0+\eta|^2)}
\kappa_N(\sqrt{2N}z_0+\xi,\sqrt{2N}z_0^*+\eta^*)
&\sim& \frac{1}{\sqrt{2\pi}}
e^{-\frac12(|\sqrt{2N}z_0+\xi|^2+|\sqrt{2N}z_0+\eta|^2
-(\sqrt{2N}z_0+\xi)^2-(\sqrt{2N}z_0^*+\eta^*)^2
)}
\nonumber\\
&&\times\mbox{erf}[(\sqrt{2N}z_0+\xi-\sqrt{2N}z_0^*-\eta^*)/\sqrt{2}] \ .
\label{kappa-zw*}
\end{eqnarray}
Here, the $N$-dependent terms do not cancel and the argument 
$X=\sqrt{N}(z_0-z_0^*)+(\xi-\eta^*)/\sqrt{2}$ of the error function becomes large in the imaginary direction
{
 (i.e. at angle $\pm\pi/2$), due to $z_0-z_0^*=\mathcal{O}(1)$. Thus we can use the following expansion \cite{NIST}
\begin{equation}
\mbox{erf}(X)= \mbox{sign}(\re(X))- \frac{1}{\sqrt{\pi} X}\exp[-X^2](1+O(1/|X|^2))\ ,
\label{erf}
\end{equation}
for $|X=R\,e^{i\Phi}|\to\infty$, with $|\Phi|<\frac{3}{4}\pi$, cf. 
\cite{note}.
}
Multiplying out the exponential factor in \eqref{kappa-zw*} we obtain this time
\begin{eqnarray}
&&\exp\left[
-\frac12\left(2N(2|z_0|^2-z_0^2-z_0^{*2}) + \sqrt{2N}\left(z_0(\xi^*+\eta^*-2\xi)+
z_0(\xi+\eta-2\eta^*)\right)
+|\xi|^2-\xi^2+|\eta|^2-\eta^{*2}
\right)
\right].
\end{eqnarray}
For the term $\pm$unity in \eqref{erf} the contribution of order $\mathcal{O}(N)$ in the exponent is again $-4Ny_0^2$, and thus it vanishes in the limit. Consequently we only obtain a contribution from the next term including  the Gaussian factor $\exp[-X^2]$. This leads to cancellations and we finally obtain 
\begin{eqnarray}
e^{-\frac12(|\sqrt{2N}z_0+\xi|^2+|\sqrt{2N}z_0+\eta|^2)}
\kappa_N(\sqrt{2N}z_0+\xi,\sqrt{2N}z_0^*+\eta^*)
&\sim& \frac{-1}{\sqrt{2 N}(z_0-z_0^*)}
K_{\rm Gin}(\xi,\eta^*)
\frac{f_{2N}(\xi)}{f_{2N}(\eta)} \ , \nonumber\\
\label{kappa-lim-Gin}
\end{eqnarray}
with $f_N$ from \eqref{f-def}. The same result applies to the complex conjugated pre-kernel, resulting into a checkerboard structure, that is every other matrix element in 
\eqref{K2x2} becomes zero in every row and column. 
In order to cancel the pre-factor in \eqref{Rk-qGin}, which would still lead to an algebraic decay to zero, and to arrive at the final answer, we need the following standard property of such a Pfaffian determinant:
\begin{equation}
\mbox{Pf}
\left[A\otimes i\sigma^y\right] = \det[A]\ .
\end{equation}
Here, $A$ is an arbitrary $N\times N$ matrix and $\sigma^y$ the second Pauli matrix, leading to such an anti-symmetric checkerboard embedding of $A$.
In summary we obtain for the limit in the bulk away from the real line to leading order 
\begin{eqnarray}
R_{k,N}^{\rm GinSE}\left(\sqrt{2N}z_0+\xi_1,\ldots,\sqrt{2N}z_0+\xi_k\right)\sim
\sqrt{2N}(z_0^*-z_0)^k
\det_{i,j=1,\ldots,k}\left[\frac{-1}{\sqrt{2 N}(z_0-z_0^*)}
K_{\rm Gin}(\xi_i,\xi_j^*)
\frac{f_{2N}(\xi_i)}{f_{2N}(\xi_j)} 
\right] .\quad
\end{eqnarray}
After cancelling all the pre-factors and using the equivalence  of kernels it agrees with the correlation functions \eqref{Rk-Gin-b} of the GinUE in the bulk of the spectrum. This finishes the bulk universality proof.
The (different) universality of the correlations functions \eqref{Rk-origin} obtained when staying close to the real axis can be shown is a similar way.

\end{appendix}
%%%%%%%%%%%%%%%%